\documentclass[manuscript,screen]{acmart}

\usepackage{booktabs}
\usepackage{arydshln}
\usepackage{longtable}
\usepackage{array}
\usepackage{graphicx}

\usepackage{multirow}
\usepackage{graphicx} 

\usepackage{tabularx}
\usepackage{makecell}

\usepackage{float}

\usepackage{placeins}
\usepackage[table]{xcolor}

\usepackage{pdflscape}
\usepackage{lscape}
\AtBeginDocument{%
  }

\setcopyright{none}
\renewcommand\footnotetextcopyrightpermission[1]{}

\begin{document}

\title{How AI Coders Discuss, Disagree, and Reach Consensus: Challenges and Opportunities for LLM-Based Qualitative Coding}

\author{Jeongyeon Kim}
\email{jeongyeon.kim@stanford.edu}
\affiliation{%
  \institution{Stanford University}
  \city{Stanford}
  \state{CA}
  \country{USA}
}

\author{John Mitchell}
\email{John.Mitchell@stanford.edu}
\affiliation{%
  \institution{Stanford University}
  \city{Stanford}
  \state{CA}
  \country{USA}
}








\renewcommand{\shortauthors}{Trovato et al.}

\begin{abstract}
  The utility of AI in multi-coder qualitative coding has been widely discussed, yet little empirical evidence exists to delineate the contexts in which it performs reliably. We address this gap by quantifying the effectiveness of multi-agent LLM coding across varied qualitative datasets, revealing key contextual and structural factors that mediate coding outcomes. We developed a literature-informed baseline pipeline that enables AI agents to independently code, debate, and reconcile disagreements. Results revealed that coding accuracy depends on factors such as codebook length, qualitative data similarity, and agent disagreement. Notably, intense and unresolved debates between agents led to higher accuracy. Our analysis showed that while LLMs emulate many human discussion behaviors, they lack adaptive responsiveness to context. From these findings, we offer design recommendations for building automated coding systems. Our open-source AI discussion dataset and methodological framework lay the groundwork for advancing the design of AI-mediated automated thematic analysis.
\end{abstract}



\maketitle

\section{Introduction}


Deductive qualitative coding is the process of applying predefined frameworks to categorize and interpret qualitative data. It is a time-consuming and labor-intensive task even for experts, demanding meticulous examination and contextual understanding of nuanced data~\cite{jiang2021supporting, kantor2024best}. The central role of qualitative coding in social science has led to a growing body of recent work on automating qualitative coding through artificial intelligence (AI) and natural language processing (NLP) techniques.
Motivated by our own interest in qualitative research with large amounts of data, we investigate AI-based coding methods modeled on human practices, such as multi-coder comparison and consensus. Our study quantifies the effectiveness of multi-agent Large Language Model (LLM) coding across varied qualitative datasets, revealing key contextual and structural factors that mediate coding outcomes.

In leveraging AI for multi-coder qualitative coding, we investigate how effectively AI can engage with the complexities of data, identifying specific conditions — when and in which contexts — AI demonstrates consistent reliability. Our analysis is intended to help researchers using our methods avoid both \emph{over-reliance} that can amplify automation bias by sidelining necessary human interpretation and \emph{under-trust} that could preclude their potential utility. Although AI tools may perform adequately in structured cases, they often falter in interpreting the nuanced subtleties of qualitative data like interview transcripts. This form of data may resist rigid categorization, requiring interpretive judgment that extends beyond mechanical rule-based classification. Even among experienced human coders, variability in interpretation is common, which is why measures like intra-rater reliability and consensus-building are indispensable. Qualitative coding is thus not merely a technical task but a deeply interpretive practice grounded in human insight.

Despite these inherent complexities of qualitative coding, recent advancements in LLMs have accelerated efforts to automate this process. They explored prompt engineering~\cite{chew2023llm, zhang2024qualitative} and incorporated human feedback loops into the automation process~\cite{zambrano2023ncoder, barany2024chatgpt}. However, these efforts remain confined to narrowly defined contexts, addressing limited or domain-specific datasets. Moreover, the prevailing focus on coding accuracy marginalizes user-centered considerations and broader system design concerns, such as transparency and interpretability. As a result, system designers are left without comprehensive, practically grounded frameworks to guide the development of automated systems that align with the needs and expectations of end users. Our research bridges these gaps by (1) investigating user needs with automated coding systems through conducting an exploratory literature review, (2) evaluating LLM capabilities across diverse datasets, and (3) proposing actionable design guidelines for future systems. 

The findings of the exploratory literature review emphasize the significance of transparency and reliability in automated coding systems. The reviewed studies demonstrate that users seek to understand the underlying coding process, including why codes are assigned and how disagreements among coders are resolved. The literature also stresses the necessity of communicating system reliability, allowing users to set realistic expectations about system outputs. Another recurrent theme across the literature is a lack of explicit optimization strategies or actionable guidance to enhance system performance, which often leads to user frustration. Drawing upon the findings from the literature review, we proposed a set of design goals aimed at improving the transparency and reliability of automated qualitative coding systems.

Following a synthesis of user needs derived from our review, we investigated the capacity of LLMs with the aim of translating identified user needs into actionable design recommendations. Based on the extracted design goals, we developed a baseline pipeline for qualitative coding, which was applied across four distinct qualitative datasets. This pipeline adopts a multi-agent architecture where AI agents simulate the collaborative coding process of human coders. The experiment results demonstrated how factors such as dataset characteristics, labeling uncertainty, and coder consensus levels influence the effectiveness of LLM-based systems. For example, coding accuracy improves when the codebook is more concise and when AI coders engage in more intense disagreements during the coding process. Additionally, by analyzing and comparing the behaviors of LLMs and human coders, we identified strategies for developers and designers to adjust LLMs for automated coding tasks. The analysis revealed that while LLMs replicate basic human interaction patterns, such as acknowledging other discussants' perspectives and proposing alternatives, they fall short in nuances, such as questioning to clarify uncertainties or drawing on personal experiences. 

These findings culminated in a set of design recommendations for creating LLM-based qualitative coding systems. The recommendations introduce strategies for designers, including guiding users of systems through data preprocessing, leveraging uncertainty and conflict as diagnostic signals for accuracy, and tuning agent discussion styles based on task ambiguity. These recommendations serve as a conceptual toolkit that helps users better understand and interact with model outputs. To advance the field, we made our baseline pipeline and the LLM agents' discussion dataset publicly available.

The contributions of our work can be summarized as below:

\begin{itemize}

\item Design goals for automated qualitative coding systems based on exploratory literature review
\item Design of baseline pipeline for LLM-based qualitative coding
\item Discussion dataset of LLM coders across four qualitative coding tasks
\item Data analysis of LLM capabilities and factors influencing coding accuracy
\item Design recommendations for LLM-based qualitative coding systems

\end{itemize}
\section{Related Work}

Even for experts, qualitative coding involves considerable cognitive and temporal investment, owing to the need for detailed scrutiny and deep contextual understanding of nuanced data~\cite{jiang2021supporting, kantor2024best}. Specifically, our work focused on deductive coding, which applies predefined theoretical frameworks to categorize and interpret qualitative data. This section addresses two dimensions in the evolution of qualitative coding systems: the automation of qualitative coding through NLP techniques and the role of human users in engaging with these automated systems.

\subsection{Automation of Qualitative Coding}

Efforts to automate qualitative coding in NLP have evolved from simple rule-based approaches to interactive machine-learning systems. Marathe and Toyama~\cite{marathe2018semi} proposed a semi-automated tool using basic NLP methods like keyword matching, while Chen et al.~\cite{chen2018using} introduced SVM models with a visualization on coder disagreement. Several studies broadened the scope of AI-assisted automated coding to encompass workflows that involve collaboration between multiple human coders~\cite{drouhard2017aeonium, gao2023coaicoder}. Building on these efforts, the advancement of LLMs has accelerated the development of automated qualitative coding techniques. For example,  Dunivin et al.~\cite{dunivin2024scalable} and Ashwin et al.~\cite{ashwin2023using} demonstrated that chain-of-thought reasoning enhances coding accuracy alongside a few-shot learning method~\cite {xiao2023supporting}. Several papers~\cite{chew2023llm, zhang2024qualitative} introduced stepwise prompt-guided coding and Liu et al.~\cite{liuassessing} highlighted the role of contextual data, such as dialogue flow and timestamps. Other studies have shown that prompt engineering can greatly enhance the accuracy of coding with LLMs. Examples include using precise codebook definitions in social science domains~\cite{halterman2024codebook}, adding coding rules informed by debates over borderline cases~\cite{meng2024exploring}, and including the reasoning behind code assignments~\cite{siiman2023opportunities}. Following these advances in prompt engineering, researchers have explored ways to make the LLM-based coding process more rigorous and systematic to improve the precision and adaptability of coding. Subramaniam et al.~\cite{subramaniam2024debategpt} proposed DebateGPT, a multi-agent system refining outputs through summarization and confidence scoring. Bryda and Sadowski~\cite{bryda2024words} iteratively adjusted prompt design to guide AI in deductive coding with clarity and consistency. Meanwhile, recognizing the need for methodological rigor and reproducibility, Tai et al.~\cite{tai2024examination} introduced a metric to measure coding consistency over repeated runs, and several other studies~\cite{bijker2024chatgpt, kirsten2024decoding} also ran multiple repeated LLM experiments to ensure reliability and robustness in their findings.

Existing research is confined to experiments on individual datasets, leaving a gap in providing a comprehensive guide for designing scalable and generalizable automated qualitative coding systems. Furthermore, as far as we are aware, there is no existing research that takes an analytical approach to how LLMs perform under varying conditions (e.g., codebook length, complexity of data to be coded). This lack of analysis of the factors influencing LLM performance makes it difficult for system developers and designers to optimize LLM performance, leaving it as an unguided trial-and-error process. Our study seeks to bridge this gap by performing a statistical analysis of LLM capabilities in qualitative coding across diverse datasets. By analyzing the relationship between dataset features and LLM performance, our work offers a foundation for understanding and optimizing LLM capabilities in qualitative coding.

\subsection{Humans' Roles in Automated Qualitative Coding}

Despite advancements in technologies for automating qualitative coding, practical adoption may remain limited if the roles and needs of human users are overlooked. In this context, prior research highlights efforts to address users' needs and challenges in automating the process of coding. Through interviews, Zhang et al.~\cite{zhang2023redefining} uncovered several user concerns regarding the use of LLMs in qualitative coding, such as a lack of transparency and limited understanding of LLMs' capabilities. Similarly, Chen et al.~\cite{chen2016challenges} highlighted users' concerns regarding the black-box nature and lack of transparency in automation models.

Jiang et al.~\cite{jiang2021supporting}, based on interviews with 17 qualitative researchers, suggested that AI-based tools should embrace ambiguity, enhance human agency, and encourage serendipitous discoveries instead of removing uncertainty. In light of these findings, further research has emphasized the active role of users in addressing limitations and refining automation processes. For instance, a few studies stressed the indispensability of user input in automated qualitative coding ~\cite{christou2023use}, such as reflecting and adapting the automated results~\cite{khan2024automating} and interpreting nuances of data~\cite{bryda2024words, carius2024artificial, morgan2023exploring}. Preserving user agency~\cite{spinoso2023qualitative, kantor2024best} and ensuring transparency~\cite{feuston2021putting, rietz2021cody, hong2022scholastic} have also been identified as key features of automated systems. These perspectives collectively highlight the importance of designing systems that not only streamline the coding process but also foster human-in-the-loop collaboration and maintain user control. In this thread, Zambrano et al.~\cite{zambrano2023ncoder} have explored the use of AI agents as collaborative coding companions, and another study~\cite{barany2024chatgpt} showed that human-AI collaboration achieves higher accuracy compared to using AI or human coders alone. Meanwhile, interactive tools like PaTAT~\cite{gebreegziabher2023patat} and CollabCoder~\cite{gao2024collabcoder} exemplified mixed-initiative approaches, allowing users to iteratively guide AI while retaining autonomy.

These approaches align with the primary goal of qualitative research, which is to foster a deeper understanding of qualitative data rather than simply speeding up the coding process. Basit et al.~\cite{basit2003manual} further underscored that the goal of qualitative coding in thematic analysis is not merely to label data but to support researchers in discovering, supporting, and confirming their theories through the data. Meanwhile, challenges of automated systems remain as Bano et al.~\cite{bano2023exploring} noted the limitations of LLMs in understanding context and nuances, cautioning against AI agents fostering a single, echo-chamber perspective. Ziems et al.~\cite{ziems2024can} evaluated LLMs' capabilities on diverse benchmarks, showing potential to enhance computational social science but highlighting the need for human oversight in the process.

Building on previous studies, our exploratory literature review investigates user requirements and expectations in automating qualitative coding. We then designed LLM experiments based on these findings. By comparing the findings of the exploratory literature review with the outcomes of the experiments, we formulate concrete design recommendations for automated qualitative coding systems. 

By addressing these gaps, our study lays the groundwork for an integration of LLMs into qualitative research practices through the investigation of user needs, statistical analysis, and design recommendations.
\section{Exploratory Literature Review}

\begin{table}[ht]
\centering
\renewcommand{\arraystretch}{1.5} 
\begin{tabular}{p{5cm} p{10cm}}

\toprule 
\multicolumn{1}{c}{\textbf{Design Goals (DG)}} & \multicolumn{1}{c}{\textbf{Supporting Literature}} \\
\midrule 

DG1. Provide reasons and evidence for how codes are assigned. & \cite{rietz2021cody, gebreegziabher2023patat, lopez2024making, overney2023sensemate, jiang2021supporting, feuston2021putting, siiman2023opportunities, kirsten2024decoding, katz2024thematic, zhang2023qualigpt, overney2024sensemate, palea2024annota, yang2025artificial, hamilton2023exploring, davison2024ethics, christou2023use, lennon2021developing, zhang2023redefining, bennis2025advancing, amani2025applying, rietz2021designing, arlinghaus2024inductive, hasija2022artificial, tschisgale2023integrating, chen2024computational, chen2018using, cameron2025breaking, johs2022explainable, costa2023qualitative, li2024comparing, piano2024qualitative, sabbaghan2024exploring, than2024updating, christou2023critical, csen2023new, perkins2024generative, wachinger2024prompts, gao2025using, brent2003feeling, tai2024examination, zikeba2025generative, de2024performing, hong2022scholastic, turobov2024using, yan2024human, nyaaba2025optimizing, qiao2025thematic, mathis2024inductive, parfenova2024automating, ashwin2023using, player2024use, tornberg2023use, bano2023ai, bano2023exploring, chew2023llm, xu2025tama, de2023can} \\

DG2. Illuminate uncertainties and interpretative variability in code assignments. & \cite{rietz2021cody, gebreegziabher2023patat, morgan2023exploring, bryda2024words, lopez2024making, overney2023sensemate, jiang2021supporting, siiman2023opportunities, kirsten2024decoding, marshall2024ethics, overney2024sensemate, mortelmans2024nvivo, palea2024annota, gao2023coaicoder, gao2023impact, hamilton2023exploring, pattyn2024value, davison2024ethics, sinha2024role, christou2023use, amani2025applying, hitch2024artificial, jalali2024integrating, nelson2020patient, paulus2024minutes, mehrad2024qualitative, christou2024thematic, rietz2021designing, spinoso2023qualitative, tschisgale2023integrating, longo2019empowering, perkins2024use, chopra2023conducting, chen2024computational, zhao2024new, chen2018using, costa2023qualitative, klemmer2024using, gao2024collabcoder, piano2024qualitative, sabbaghan2024exploring, goyanes2024thematic, than2024updating, wheeler2025use, christou2023use, kantor2024best, csen2023new, perkins2024generative, li2021qualitative, dunivin2024scalable, dengel2023qualitative, zade2018conceptualizing, tai2024examination, christou2025looking, zikeba2025generative, anis2023efficient, de2024performing, dai2023llm, yan2024human, nyaaba2025optimizing, nguyen2025narrative, lee2024harnessing, friedman2024should, hayes2025conversing, eschrich2024framework, qiao2025thematic, de2024reflections, atasautomating, mathis2024inductive, parfenova2024automating, carius2024artificial, rasheed2024can, ritschlassessing, smirnov2025enhancing, de2024applications, nejjar2025llms, khan2024automating, omizo2024automating, bano2024large, bano2023ai, roberts2024artificial, chew2023llm, rao2024quallm, de2023can, nelson2021future, nguyen2024chatgpt} \\

DG3. Enable users to calibrate expectations for coding accuracy. & \cite{gebreegziabher2023patat, marshall2024ethics, zhang2023qualigpt, overney2024sensemate, palea2024annota, gao2023coaicoder, gao2023impact, christou2023use, zhang2023redefining, xiao2023supporting, hasija2022artificial, johs2022explainable, klemmer2024using, zade2018conceptualizing, hong2022scholastic, bano2024large} \\

DG4. Facilitate users' understanding of how data characteristics affect coding accuracy. & \cite{rietz2021cody, bryda2024words, overney2023sensemate, feuston2021putting, siiman2023opportunities, kirsten2024decoding, marshall2024ethics, prescott2024comparing, overney2024sensemate, mortelmans2024nvivo, gao2023impact, yang2025artificial, hamilton2023exploring, amani2025applying, bijker2024chatgpt, liu2025qualitative, chubb2023me, tschisgale2023integrating, gao2024collabcoder, giri2024rapid, mesec2023language, wachinger2024prompts, dunivin2024scalable, brondani2024artificial, anis2023efficient, guetterman2018augmenting, hayes2025conversing, atasautomating, de2024applications, ashwin2023using, player2024use, bano2023exploring, chew2023llm} \\

\bottomrule 
\end{tabular}
\caption{An overview of the design goals for automated qualitative coding systems and literature that informed their formulation.}
\label{tab:design_goals_and_lit_review}
\end{table}

The exploratory literature review explores user needs and expectations for automated qualitative coding systems, aiming to derive design goals that inform the development of a baseline LLM-based coding pipeline. We adopted the PRISMA statement~\cite{moher2009preferred} for review.

\subsection{Review Methodology}

\subsubsection{Identification Process}

For the identification of relevant literature, we conducted a database search using the following keywords: (“AI” OR “artificial intelligence” OR “LLMs” OR “large language models” OR “automated”) AND (“qualitative coding” OR “qualitative analysis” OR “thematic analysis” OR “thematic coding”). This search yielded an initial set of 988 articles, which were sorted and selected based on their relevance to the specified keywords.

\subsubsection{Screening Process}

One of the authors reviewed the titles and abstracts of all retrieved papers. Papers were excluded if their focus was unrelated to automated qualitative coding or if they were not written in English. For example, papers using the term "coding" in the context of computer programming rather than thematic analysis were removed. This screening procedure resulted in a final set of 180 relevant papers.

\subsubsection{Assessing Eligibility}

To determine the eligibility of relevant literature, one of our authors read full articles of 180 papers. Through an iterative process of reading and analysis, the author developed the following exclusion criteria:

\begin{itemize}

\item Papers were excluded if the primary focus was not on automated qualitative coding. For instance, studies that employed qualitative methods such as interviews or surveys as part of broader research themes (e.g., AI in Education: A Qualitative Analysis) were omitted.
\item Papers that discussed solely human-driven coding practices without reference to automated systems were excluded (e.g., Analyzing Qualitative Data via Collaborative Coding: A Practical Guide).
\item Papers were excluded if they lacked explicit discussions of user expectations or design considerations in automated qualitative coding systems. This includes works that merely presented use cases of technologies or technical architectures without addressing user-side design requirements.

\end{itemize}

The exclusion criteria were refined throughout the iterative review process. As a result of this eligibility evaluation, a final corpus of 118 papers was selected for analysis. The complete list of reviewed papers is provided in the Supplementary Materials.

\subsection{Analysis}

Given that our aim was to inform design goals for our baseline pipeline of LLM-based qualitative coding, we conducted an exploratory review rather than a comprehensive,  systematic literature review. One of our authors generated preliminary codes through a full reading of the corpus, followed by a second round of analysis to finalize the codebook. We then invited a qualitative research expert with a PhD in sociology to independently assess the codebook against the full corpus. 

\subsection{Review Results}

We present the design goals identified through the review of prior literature. Table~\ref{tab:design_goals_and_lit_review} provides an overview of these goals and their supporting studies.

\subsubsection{Transparency}

Coding automation should help users make informed decisions by presenting outputs that are transparent in their reasoning and limitations, including an explanation of how results are derived and any associated uncertainties.

\textbf{DG1. Provide reasons and evidence for how codes are assigned.}
Prior work emphasizes that automated coding systems should offer explanations for their coding outputs \cite{rietz2021cody, zhang2023qualigpt} to support user understanding of AI reasoning \cite{gebreegziabher2023patat, christou2023use}. To reflect this, we prompted the LLM to include justifications when generating codes.

\textbf{DG2. Illuminate uncertainties and interpretative variability in code assignments.}
Literature highlights the importance of exposing uncertainties in AI-generated codes, often represented through confidence scores \cite{li2021qualitative, gebreegziabher2023patat, klemmer2024using}. Furthermore, prior work underscores interpretive variability, across both human and AI coders \cite{siiman2023opportunities, dai2023llm}. Based on this, we used two LLM coders to perform coding independently and resolve disagreements through discussion, mimicking human coding processes. We instructed the models to estimate confidence level, allowing LLM coders to assign an ‘Undecidable’ label when uncertain.

\subsubsection{Reliability}

Automated coding systems should enable users to better understand the system's capabilities and develop strategies to optimize performance.

\textbf{DG3. Enable users to calibrate expectations for coding accuracy.}
Users often struggle when they cannot align expectations with system performance~\cite{christou2023use, palea2024annota}. In line with established approaches~\cite{bano2023exploring, zade2018conceptualizing}, we used coder disagreements during discussions as a proxy for evaluating automated coding accuracy. 

\textbf{DG4. Facilitate users' understanding of how data characteristics affect coding accuracy.}
Literature suggests that factors such as codebook complexity~\cite{bryda2024words}, the abstractness of language~\cite{prescott2024comparing, overney2023sensemate}, and data similarity~\cite{amani2025applying, bijker2024chatgpt} affect coding accuracy. Our study investigated how variations in data and codebook (e.g., length, similarity, and specialization) impact coding reliability.
\section{Large Language Model Experiments and Data Analysis}

In this section, we report the pipeline and results of LLM experiments designed based on the design goals. Our goal is to examine the capabilities of LLMs in qualitative coding by evaluating their performance across different datasets.

\subsection{Datasets}

We conducted experiments using LLMs across four distinct domains: education, law, sociology, and medicine, utilizing datasets from prior studies. The datasets include a dialogue corpus in learning~\cite{wachsmuth2022mama}, cases from the European Court of Human Rights~\cite{chalkidis2021paragraph}, open-ended interview responses~\cite{sanfilippo2020shuffle}, and medical paper abstracts~\cite{schopf2022evaluating}. To standardize dataset sizes, we randomly selected five labels from each corpus and sampled 500 instances per label. Meanwhile, we have released the prompt design, coding pipeline, and LLM-generated discussions as open-source materials\footnote{\href{https://anonymous.4open.science/r/qual-coding-llm-experiment-1DA3/README.md}{https://anonymous.4open.science/r/qual-coding-llm-experiment-1DA3/README.md}}.

\subsection{Design of Coding Pipeline}

Based on the exploratory literature review, we designed an automated coding pipeline powered by AI agents. The process mirrors human coding in that the AI coders first independently code the entire dataset and two AI coders engage in discussions about conflicting cases (\textbf{DG2}). In the discussion process, they aim to resolve conflicts and derive new labeling rules based on the discussions. 

An important note is that, before any of these steps, an initial phase of extracting excerpts to be coded was conducted. This data preprocessing step aimed to standardize the units of analysis. In this step, the LLM was supplied with the codebook and the data to be coded, and it was instructed to extract data excerpts likely to correspond to the codes in the codebook. As a result, only the extracted excerpts, rather than the entire qualitative dataset, were used for coding. This approach aligns with methodologies discussed in recent studies, which emphasize the importance of preprocessing to improve the reliability of qualitative coding by ensuring only pertinent excerpts are subjected to analysis~\cite{gao2023impact, than2024updating, schwitter2025using}.

The prompts used in the coding process were structured as follows. AI coders were instructed to justify their coding decisions (\textbf{DG1}) and had three possible response options: the code is present in the data, absent from the data, or undecidable (\textbf{DG2}). We grounded the prompt design in prior work, adhering to principles of constructive discussions~\cite{scannapieco1997formal, deutsch2011handbook, summers1950debate} and efficient AI-agent dialogue frameworks~\cite{li2024more, wang2024rethinking, chan2023chateval, smitshould}. To reduce ambiguity in the coding task for AI models, we followed established conventions in the literature by including only one code per coding cycle and repeating the process for each code individually. As the discussion progressed through multiple rounds and turns, LLM coders were encouraged to formulate their answers by considering both their previous reasoning and the other coder’s perspectives. If disagreements arose initially, but consensus was achieved through rounds of discussions, the coders collaboratively developed a new labeling rule to resolve similar disputes in the future. Each back-and-forth exchange constituted one round of discussion. If no agreement was reached after three rounds, the case was marked as “disagreed,” and the process moved on to the next case. To avoid ordering bias, we adhered to methodologies established in reasoning and QA frameworks for AI-agent discussions~\cite{du2023improving}. We utilized OpenAI’s ChatGPT with the ‘gpt-4o-mini’ model for this process. The temperature settings were 0 for coding and 0.7 for discussions. These configurations align with previous studies on qualitative coding, but future researchers may adjust them to meet their objectives. After completing the discussion phase, both AI coders re-coded the entire dataset using shared knowledge derived from discussions and newly established labeling rules. At each coding round, we calculated inter-rater reliability (IRR) using Cohen’s Kappa.

\subsection{Evaluation Method}

Following the pipeline design, we addressed data imbalance using an undersampling technique~\cite{liu2008exploratory} and computed F1 scores to assess coding accuracy. We then examined the relationship between coding accuracy and factors related to datasets and AI-agent discussions (\textbf{DG4}). These factors included dataset characteristics, such as length of codebooks, semantic similarity of content within codebooks, length of data excerpt to be coded, semantic similarity of content within excerpts, and disparity of specialization degree between codebooks and excerpts. Additionally, labeling dynamics, such as labeling uncertainty, degree of consensus and controversy between coders, were also considered (\textbf{DG3}), drawing upon the literature detailed in Table~\ref{tab:analysis_1_dataset_characteristics}, \ref{tab:analysis_2_labeling_uncertainty}, \ref{tab:analysis_3_coder_consensus}, and \ref{tab:analysis_4_discussion_mode}.

We quantify corpus’ semantic proximity using cosine similarity, one of the most common similarity metrics, enabling a scale-invariant comparison of textual representations~\cite{reimers2019sentence}. Meanwhile, to measure the degree of specialization of corpus, we adopted a frequency-based approach that follows practice in corpus linguistics and psycholinguistics, lower reference-corpus frequency indicates higher lexical difficulty~\cite{breland1996word, chen2016characterizing}. Specifically, we used the Corpus of Contemporary American English (COCA) to compute a Word Difficulty Score (WDS) for each word, defined by the inverse frequency of the word~\cite{tang2023assessing, corpuscorpus}. 

For statistical analysis, we adopted Mixed-Effects Models, an approach designed to handle both fixed and random effects, making it particularly suited for grouped or hierarchical data~\cite{schmettow2015tutorial, agresti2000random, luger2014robust}. Given the grouping structure of our data — comprising four datasets with five labels in each dataset — we modeled these groupings as controlled random effects. Model appropriateness was verified through assessments of VIF, ICC, and residual normality. In addition, we conducted a qualitative evaluation of AI coder discussion patterns. Using the Thomas-Kilmann Conflict Mode Instrument~\cite{thomas2008thomas} and frameworks for discursive moves~\cite{walton2008argumentation, zakharov2021discourse}, we compared the behaviors of human and AI discussions. One of our authors iteratively coded the patterns in AI-generated discussions and resolved disagreements.

\subsection{Statistical Analysis Results}

Across all datasets and labels, the IRRs between two AI coders exceeded 0.85. The F1 scores ranged from 0.31 to 0.89 depending on dataset and label, with an average of 0.68 and a standard deviation of 0.16. From this data, we specifically analyzed cases where either one of the two AI coders assigned the ‘Undecidable’ label or where the two coders provided conflicting results, estimating the accuracy of these cases. The average accuracy was 0.60 with a standard deviation of 0.25, highlighting the tradeoff between coverage and accuracy. Lower confidence levels and consensus degrees resulted in reduced accuracy. Full evaluation details are available in the Supplementary Material.

The Mixed Effects Model analysis revealed how coding accuracy is influenced by data characteristics and coding dynamics, as summarized in Table~\ref{tab:analysis_1_dataset_characteristics}, \ref{tab:analysis_2_labeling_uncertainty}, \ref{tab:analysis_3_coder_consensus}, and \ref{tab:analysis_4_discussion_mode}. Detailed statistics, including minimum, maximum, average, and standard deviations for the analysis, are available in the Supplementary Materials.

\subsubsection{Relationship between Coding Accuracy and Dataset Characteristics}

\begin{longtable}{
  >{\raggedright\arraybackslash}p{0.9cm}
  >{\raggedright\arraybackslash}p{1.3cm}
  >{\raggedright\arraybackslash}p{1.5cm}
  >{\raggedright\arraybackslash}p{2.5cm}
  >{\raggedright\arraybackslash}p{3.3cm}
  >{\raggedright\arraybackslash}p{4cm}
}

\toprule
\textbf{Metric} & 
\textbf{Dataset Characteristic} & 
\textbf{P-value (T-statistic, Effect Size)} & 
\textbf{Relationship} & 
\textbf{Reason and Interpretation} & 
\textbf{Relevant Literature} \\
\midrule
\endfirsthead

\toprule
\textbf{Metric} & 
\textbf{Dataset Characteristic} & 
\textbf{P-value (T-statistic, Effect Size)} & 
\textbf{Relationship} & 
\textbf{Reason and Interpretation} & 
\textbf{Relevant Literature} \\
\midrule
\endhead

\multirow{5}{=}{Accuracy of Initial Coding} 

& Length of Codebook 
& \cellcolor{red!15}< 0.0001 **** (-11.702, -0.287) 
& The shorter the codebook, the higher the initial accuracy of automated coding. 
& LLMs' ability to retrieve and retain relevant information from the input declines as the context length increases. 
& - The performance of LLMs degrades when identifying relevant information within lengthy contexts~\cite{liu2024lost, li2024long}. \newline - As context length increases, the model's ability to focus on the most relevant information can decrease, leading to a ``dilution'' of attention across the input~\cite{xu2023retrieval}. \newline - To create a good codebook in qualitative data analysis, ensure it is clear and concise~\cite{barany2024chatgpt, ritchie2022development, decuir2011developing}. \\

& Cosine Similarity of Codebook 
& \cellcolor{red!15}< 0.0001 **** (-11.738, -0.276) 
& The less similar the semantics of words in each codebook, the higher the initial accuracy of automated coding. 
& Describing code's meaning in varied language provides multiple perspectives and contexts for the code compared to using overly specific or narrow language.
& Not Applicable \\

& Length of Excerpt 
& \cellcolor{red!15}< 0.0001 **** (-9.249, -0.438) 
& The shorter the excerpt to code, the higher the initial accuracy of automated coding. 
& Concise data contains less noise from extraneous details, making it easier for AI to focus on the main idea. 
& - The performance of LLMs degrades when identifying relevant information within lengthy contexts~\cite{liu2024lost, li2024long}. \newline - The long and unsegmented text segments made coding decisions more complex~\cite{hruschka2004reliability}. \\

& Cosine Similarity of Excerpt 
& \cellcolor{green!15}< 0.0001 **** (7.372, 0.189) 
& The more similar the semantics of words in excerpt to code, the higher the initial accuracy of automated coding. 
& Limited or consistent topics of data to code make it easier for the AI to identify patterns and relationships within the data. 
& - LLMs perform better in cases where fewer, more concentrated topics allow the model to capture relevant information easily~\cite{doi2024comprehensive}. \\

& Gap of Degree of Specialization Between Codebook and Excerpt 
& \cellcolor{red!15}< 0.0001 **** (-7.629, -0.331) 
& The smaller the gap of degree of specialization between codebook and excerpt to code, the higher the initial accuracy of automated coding. 
& Text alignment in NLP involves mapping corresponding segments between two texts by measuring their similarity and connecting the most similar parts. Qualitative coding requires similar alignment between the codebook and data. A gap in technical terminology use between the two lowers the similarity score, which can reduce coding accuracy.
& - Text alignment relies on the assessment of similarity to establish accurate correspondences across different text segments~\cite{kay1993text, zha2023text}. \newline - Significant terminological differences between sources — such as technical versus lay language — impair the model's ability to form coherent personas. Lacking shared features like vocabulary and style, the model struggles to maintain cross-domain accuracy and truthfulness~\cite{joshi2023personas}. \\

\bottomrule

\caption{Dependency on Dataset Characteristics. This table summarizes how different dataset characteristics affect the accuracy of AI agents in initial coding. The summary of the analysis, conducted using a Mixed Effects Model, includes statistical significance, reasons and interpretations of the findings, and connections to relevant literature. Green cell indicates positive, red negative, and gray no relationship with coding accuracy. *: p<0.05, **: p<0.01, ***: p<0.001, ****: p<0.0001.}
\label{tab:analysis_1_dataset_characteristics} \\
\end{longtable}

Table~\ref{tab:analysis_1_dataset_characteristics} illustrates the relationship between dataset characteristics and coding accuracy. Lengthier codebooks, higher cosine similarity within the codebook, lengthier excerpts, and greater gaps in specialization between codebooks and excerpts all negatively affect accuracy. In contrast, higher cosine similarity within the excerpt positively impacts coding accuracy.

We observed that concise codebooks improve AI coders' labeling accuracy, as LLMs struggle to retrieve and retain information with extended input contexts (Table~\ref{tab:analysis_1_dataset_characteristics}, Row 1). This finding aligns with previous studies emphasizing the efficiency of concise inputs for LLMs~\cite{liu2024lost, li2024long, xu2023retrieval, barany2024chatgpt, ritchie2022development, decuir2011developing}. One should be cautious in interpreting this finding, as straightforward codes often require less detailed descriptions, whereas complex codes necessitate nuanced, lengthy explanations. This means that the observed high accuracy with concise inputs may partly reflect the complexity of the code content itself, rather than the conciseness alone. To account for this, we incorporated specialized terminology and Word Sense Disambiguation (WSD) to minimize the complexity's influence. This further analysis mitigated the potential confounding effects of complexity, which our analysis found to be statistically insignificant. The same pattern was also observed in the data to be coded (Table~\ref{tab:analysis_1_dataset_characteristics}, Row 3), where AI coders performed more accurately with concise data. 

The analysis revealed opposing effects of cosine similarity within the codebook (Table~\ref{tab:analysis_1_dataset_characteristics}, Row 2) versus the data to be coded (Table~\ref{tab:analysis_1_dataset_characteristics}, Row 4). Codebooks employing diverse languages improved accuracy, as variety aids in contextual understanding of the code. Conversely, consistency in topics within the data to be coded improved accuracy, as fewer topics made it easier for the model to capture relevant information easily~\cite{doi2024comprehensive}. To summarize, coding accuracy improves when codebooks employ diverse language while the target data remains consistent. Since the coding process requires determining whether individual data points fall within codebook categories, this difference in linguistic breadth — varied vocabulary in the codebook and the homogeneity of languages in data — leads to more accurate coding results. It suggests that designing a corpus requires balancing diversity with specificity, depending on the role of each corpus. This balance highlights the complementary nature of broad comprehension (for codebooks) and targeted application (for data). 

Additionally, drawing parallels from text alignment tasks in NLP domain, which involves mapping corresponding text segments across sources, we explored the impact of similarity alignment between codebooks and data. The result showed that significant disparities in technical terminology diminished similarity alignment, impairing coding accuracy (Table~\ref{tab:analysis_1_dataset_characteristics}, Row 5). This result corroborates existing research showing that differences in terminology reduce LLM efficacy~\cite{kay1993text, zha2023text, joshi2023personas}.

\subsubsection{Relationship between Coding Accuracy Increase and Labeling Uncertainty}

\begin{longtable}{
  >{\raggedright\arraybackslash}p{1.5cm}
  >{\raggedright\arraybackslash}p{1.8cm}
  >{\raggedright\arraybackslash}p{1.5cm}
  >{\raggedright\arraybackslash}p{2.9cm}
  >{\raggedright\arraybackslash}p{3.3cm}
  >{\raggedright\arraybackslash}p{3.2cm}
}

\toprule
\textbf{Metric} & 
\textbf{Labeling Uncertainty} & 
\textbf{P-value (T-statistic, Effect Size)} & 
\textbf{Relationship} & 
\textbf{Reason and Interpretation} & 
\textbf{Relevant Literature} \\
\midrule
\endfirsthead

\toprule
\textbf{Metric} & 
\textbf{Labeling Uncertainty} & 
\textbf{P-value (T-statistic, Effect Size)} & 
\textbf{Relationship} & 
\textbf{Reason and Interpretation} & 
\textbf{Relevant Literature} \\
\midrule
\endhead

\multirow{2}{=}{Accuracy Increase via Discussions} 

& Number of Undecidable Labels in Initial Coding & \cellcolor{green!15}< 0.01 ** (3.274, 0.123) & The higher the number of ‘Undecidable’ labels before discussion, the greater the accuracy improvement through AI agent discussions. & Dialogue process naturally clarifies uncertainties and builds a precise, shared understanding, thereby improving coding accuracy. & Language model performance is enhanced by emphasizing the uncertain data during the training process~\cite{yoo2019learning, settles2009active}. \\
& Number of Undecidable Labels in Discussion-Prompted Coding & \cellcolor{red!15}< 0.0001 **** (-14.476, -0.369) & The lower the number of ‘Undecidable’ labels after discussion, the greater the accuracy improvement through AI agent discussions. & Low frequency of ‘Undecidable’ labels serves as an indicator of the effect of AI agent discussions. & Low confidence and agreement among coders are indicators of poor performance~\cite{ganji2018ease}. \\
\bottomrule
\caption{Dependency on Labeling Uncertainty. This table summarizes how labeling uncertainty among AI agents affects the accuracy improvements achieved through AI agent discussions. The summary of the analysis, conducted using a Mixed Effects Model, includes statistical significance, reasons and interpretations of the findings, and connections to relevant literature. Green cell indicates positive, red negative, and gray no relationship with coding accuracy. *: p<0.05, **: p<0.01, ***: p<0.001, ****: p<0.0001.                               }
\label{tab:analysis_2_labeling_uncertainty} \\
\end{longtable}

Table~\ref{tab:analysis_2_labeling_uncertainty} presents the relationship between coding accuracy improvement through discussions and labeling uncertainty, measured by the frequency of ‘Undecidable’ labels assigned by AI coders. The results reveal a positive relationship between coding accuracy increase via discussions and the ratio of ‘Undecidable’ labels during initial coding before discussions. In contrast, after discussion-prompted coding (i.e., coding performed by LLM coders using a prompt that included the two coders’ discussion process and the new labeling rule derived from it), this trend reverses, showing a negative relationship. The positive relationship suggests that greater ambiguity in the initial coding task enhances the effectiveness of AI agent discussions, as the dialogue process naturally clarifies uncertainties and builds a shared, precise understanding (Table~\ref{tab:analysis_2_labeling_uncertainty}, Row 1). These findings corroborate the studies that revealed the positive relationship between clarification of labeling uncertainty and language model performance~\cite{yoo2019learning, settles2009active}. On the other hand, the frequency of ‘Undecidable’ labels persisted following the discussions serves as an indicator of the effect of AI agent discussions — more occurrences indicate a small improvement in accuracy through discussions (Table~\ref{tab:analysis_2_labeling_uncertainty}, Row 2). This result parallels findings from human coding studies, where low coder confidence levels often signal poor coding performance~\cite{ganji2018ease}.

\subsubsection{Relationship between Coding Accuracy Increase and Consensus Level of AI Coders}

\begin{longtable}{
  >{\raggedright\arraybackslash}p{1.5cm}
  >{\raggedright\arraybackslash}p{1.8cm}
  >{\raggedright\arraybackslash}p{1.5cm}
  >{\raggedright\arraybackslash}p{2.9cm}
  >{\raggedright\arraybackslash}p{3.3cm}
  >{\raggedright\arraybackslash}p{3.2cm}
}

\toprule
\textbf{Metric} & 
\textbf{Consensus among Coders} & 
\textbf{P-value (T-statistic, Effect Size)} & 
\textbf{Relationship} & 
\textbf{Reason and Interpretation} & 
\textbf{Relevant Literature} \\
\midrule
\endfirsthead

\toprule
\textbf{Metric} & 
\textbf{Consensus among Coders} & 
\textbf{P-value (T-statistic, Effect Size)} & 
\textbf{Relationship} & 
\textbf{Reason and Interpretation} & 
\textbf{Relevant Literature} \\
\midrule
\endhead

\multirow{2}{=}{Accuracy Increase via Discussions} 
& Number of Conflicting Labels in Initial Coding & \cellcolor{gray!15}0.059 (-1.888, -0.037) & No relationship between the number of conflicting labels before discussion and the accuracy improvement through AI agent discussions. & There is no evidence that the number of conflicting labels before discussion serves as an indicator of the accuracy improvement through AI agent discussions. & Not Applicable \\
& Number of Conflicting Labels in Discussion-Prompted Coding & \cellcolor{red!15}< 0.0001 **** (-6.861, -0.147) & The lower the number of conflicting labels after discussion, the greater the accuracy improvement through AI agent discussions. & Low frequency of conflicting labels serves as an indicator of the effect of AI agent discussions. & Low confidence and agreement among coders are indicators of poor performance~\cite{ganji2018ease}. \\
\bottomrule
\caption{Dependency on Consensus Among Coders. This table summarizes how consensus among AI agents during coding affects the accuracy improvements achieved through AI agent discussions. The summary of the analysis, conducted using a Mixed Effects Model, includes statistical significance, reasons and interpretations of the findings, and connections to relevant literature. Green cell indicates positive, red negative, and gray no relationship with coding accuracy. *: p<0.05, **: p<0.01, ***: p<0.001, ****: p<0.0001.}
\label{tab:analysis_3_coder_consensus} \\
\end{longtable}

As shown in Table~\ref{tab:analysis_3_coder_consensus}, coding accuracy improvement varies depending on the agreement levels between AI coders. Our analysis revealed that the number of conflicting labels before discussion did not influence the accuracy gains achieved through AI agent discussions (Table~\ref{tab:analysis_3_coder_consensus}, Row 1). In contrast, a low frequency of conflicting labels after discussion is indicative of the positive impact of AI agent discussions (Table~\ref{tab:analysis_3_coder_consensus}, Row 2). This is consistent with existing literature that identifies low agreement among human coders as a marker of poor qualitative coding accuracy~\cite{ganji2018ease}.

\subsubsection{Relationship between Coding Accuracy and Controversy Level of AI Coders}

\begin{longtable}{ 
>{\raggedright\arraybackslash}p{1.5cm} 
>{\raggedright\arraybackslash}p{1.8cm} 
>{\raggedright\arraybackslash}p{1.5cm} 
>{\raggedright\arraybackslash}p{2.9cm} 
>{\raggedright\arraybackslash}p{3.3cm} 
>{\raggedright\arraybackslash}p{3.2cm}
}

\toprule
\textbf{Metric} & \textbf{Controversy in Discussions} & \textbf{P-value (T-statistic, Effect Size)} & \textbf{Relationship} & \textbf{Reason and Interpretation} & \textbf{Relevant Literature} \\
\midrule
\endfirsthead

\toprule
\textbf{Metric} & \textbf{Controversy in Discussions} & \textbf{P-value (T-statistic, Effect Size)} & \textbf{Relationship} & \textbf{Reason and Interpretation} & \textbf{Relevant Literature} \\
\midrule
\endhead

\multirow{6}{=}{Accuracy of Discussion-Prompted Coding} & Number of Turns in Discussions & \cellcolor{green!15}< 0.0001 **** (7.997, 0.168) & The more controversial the discussions, the higher the accuracy of discussion-prompted coding. & Discussions about controversial cases reveal nuanced details that clarify the scope of the code's application. & To create a good codebook in qualitative data analysis, include clear inclusion and exclusion criteria for each code~\cite{macqueen1998codebook, ritchie2022development}. \\

& Ratio of Resolved Conflicts & \cellcolor{red!15}< 0.0001 **** (-6.300, -0.139) & The lower the agreement ratio in the discussions, the higher the accuracy of discussion-prompted coding. & Debates with disagreements reveal nuanced details that clarify the scope of the code's application. & \\

& Ratio of Correctly Resolved Conflicts & \cellcolor{red!15}< 0.0001 **** (-9.720, -0.169) & The lower the number of correctly resolved conflicting cases via discussions, the greater the accuracy of discussion-prompted coding. & Small number of correctly resolved conflicting cases indicates that the AI agent discussions involved a high proportion of challenging and borderline cases. These discussions about borderline cases reveal nuanced details that clarify the scope of the code's application. & Not Applicable \\

& Ratio of Competing Modes in Discussions & \cellcolor{gray!15}< 0.01 ** (2.789, 0.054) & No relationship between the ratio of competing modes of AI agents in discussions and the accuracy of discussion-prompted coding. & There is no evidence that the ratio of competing modes of AI agents in discussions serves as an indicator of the accuracy of discussion-prompted coding. & Not Applicable \\

& Ratio of Compromising Modes in Discussions & \cellcolor{red!15}< 0.0001 **** (-4.251, -0.066) & The less compromising the discussions, the higher the accuracy of discussion-prompted coding. & Discussions about controversial cases, which provokes less compromising discussions among AI agents, reveal nuanced details that clarify the scope of the code's application. & - Addressing conflicts in a collaborative, not competitive manner turns disagreements into chances for improved decision-making~\cite{tjosvold2019managing}.\newline
- Collaborating mode is desired in coding discussion, seeking for consensual decision~\cite{thomas2008thomas}. \\

& Ratio of Collaborating Modes in Discussions & \cellcolor{red!15}< 0.0001 **** (-8.420, -0.255) & The less collaborating the discussions, the higher the accuracy of discussion-prompted coding. & Discussions about controversial cases, which provokes less collaborating discussions among AI agents, reveal nuanced details that clarify the scope of the code's application. & \\
\bottomrule
\caption{Dependency on Controversy in Discussions. This table summarizes how the level of controversy in AI agent discussions affects the accuracy of AI agents when prompted with the discussion process. The summary of the analysis, conducted using a Mixed Effects Model, includes statistical significance, reasons and interpretations of the findings, and connections to relevant literature. Green cell indicates positive, red negative, and gray no relationship with coding accuracy. *: p<0.05, **: p<0.01, ***: p<0.001, ****: p<0.0001. Note that the term ``competing/compromising/collaborating mode'' relates to the content’s characteristics, not the discussion styles of the AI agents, as we did not adjust the AI agents' discussion personas.}
\label{tab:analysis_4_discussion_mode} \\
\end{longtable}

Table~\ref{tab:analysis_4_discussion_mode} illustrates the relationship between accuracy of discussion-prompted coding and the degree of controversy between AI coders. Accuracy was positively related to the number of discussion turns, with contentious disagreements extending debates and leading to an increase in turns. In contrast, accuracy showed a negative relationship with the ratio of resolved conflicts, correctly resolved conflicts, as well as discussions conducted in compromising and collaborating modes. There was no significant relationship observed with the ratio of competing modes.

The high coding accuracy observed in controversial discussions can be attributed to the nuanced understanding generated during debates. Lengthier debates allowed for clarification of the inclusion and exclusion criteria for each code, aligning with principles for designing effective codebooks by exposing clear inclusion and exclusion criteria~\cite{macqueen1998codebook, ritchie2022development} (Table~\ref{tab:analysis_4_discussion_mode}, Row 1). Similarly, these effects were also prominent in discussions without consensus (Table~\ref{tab:analysis_4_discussion_mode}, Row 2). Interestingly, a smaller number of correctly resolved conflicts correlated with higher accuracy, indicating that debates about challenging or borderline cases uncovered subtle details that revealed the scope of code application (Table~\ref{tab:analysis_4_discussion_mode}, Row 3). This underscores that the value of engaging in nuanced debates may outweigh the importance of achieving correct resolutions, as the process itself drives a deeper understanding of ambiguous cases.

Contrary to findings from human discussions, where collaborative modes are typically desired, which provide emotional support and psychological safety~\cite{tjosvold2019managing, thomas2008thomas}, less compromise and collaboration among AI coders produced better results (Table~\ref{tab:analysis_4_discussion_mode}, Rows 5 and 6). This discrepancy can be explained by the nature of LLM agents, which operate free of emotional and social dynamics. Competitive discussions, which might induce stress in humans, function as optimization tools for LLMs as they can drive more rigorous refinement of reasoning. 

Additionally, LLMs typically share the same foundational architecture and training data, thereby sometimes generating similar opinions among agents. Accordingly, collaborative modes can lead to overly convergent answers and reinforce biases~\cite{anderson2024homogenization, chuang2023simulating, bano2023exploring}.

\subsection{Qualitative Observation of AI-Agent Discussions}

\begin{longtable}{
  >{\raggedright\arraybackslash}p{2.1cm}
  >{\raggedright\arraybackslash}p{2.5cm}
  >{\raggedright\arraybackslash}p{2.5cm}
  >{\raggedright\arraybackslash}p{5.5cm}
}

\toprule
\textbf{Discursive Move of LLMs} & 
\textbf{Description} & 
\textbf{Discursive Move of Humans in Literature} & 
\textbf{Examples from AI Agent Discussions} \\
\midrule
\endfirsthead

\toprule
\textbf{Discursive Move of LLMs} & 
\textbf{Description} & 
\textbf{Discursive Move of Humans in Literature} & 
\textbf{Examples from AI Agent Discussions} \\
\midrule
\endhead

\textbf{Stance Maintenance} & 
Reaffirming a viewpoint to emphasize commitment to a particular perspective. & 
Clarification~\cite{grice19901975, bakhtin2010dialogic, chan2023chateval} & 
``I \textbf{maintain} that it should be labeled as 1 for the presence of a check question.''
\newline
``However, I \textbf{still believe} it primarily serves a rhetorical purpose rather than acting as a direct check question.''
\newline
``I \textbf{still assert} that the claim fundamentally advocates for the inclusion of minority rights within a multicultural framework.''
\\

\textbf{Acknowledgment Before Disagreement} & 
Acknowledging the opposing coder's viewpoint before clearly expressing disagreement. & 
Tolerant Divergence~\cite{felton2009deliberation, lu2011collaborative, angouri2010you, cheng2017anyone}, Partial Concurrence~\cite{felton2009deliberation, lu2011collaborative, angouri2010you, cheng2017anyone} & 
``I \textbf{understand} Coder 2's concerns about the nuances of the phrase ... \textbf{However}, I \textbf{still assert} that this utterance functions as a check question.''
\newline
``I \textbf{understand} your perspective, \textbf{but} I would \textbf{argue} that the mere presence of poor conditions alone does not meet the threshold of a violation of the right to life as defined in Article 2.''
\newline
``\textbf{While} I \textbf{acknowledge} Coder 2's point regarding `severe cardiorespiratory collapse,' it is \textbf{essential to clarify} that this term alone does not constitute a cardiovascular disease as defined by the provided description, which focuses on chronic or specific conditions related to the circulatory system.'' \\

\textbf{Softening Language} & 
Using softer and cooperative language to maintain a positive and constructive atmosphere. & 
Affirming Communication~\cite{felton2009deliberation, lu2011collaborative, angouri2010you, cheng2017anyone} & 
``I \textbf{appreciate} Coder 2’s reevaluation and the acknowledgment of ambiguity.''
\newline
``Coder 1's assessment \textbf{correctly identifies} it as a form of agreement that confirms the sentiment previously discussed.'' \\

\textbf{Reassessing Evidence} & 
Revisiting the description of code, often mentioning specific lines to reinforce their position or make it more convincing. & 
Referenced Argumentation~\cite{felton2009deliberation, lu2011collaborative, angouri2010you, cheng2017anyone} & 
``According to the \textbf{requirements of D01}, the utterance must specifically ask a check question.''
\newline
``\textbf{The right to life [which is one of the codes in the codebook]} encompasses not only protection from direct life-threatening actions but also the provision of a safe and healthy environment.''
\newline
``The essence of \textbf{Article 5 [which is one of the codes in the codebook]} is focused on actual deprivation of liberty through arrest or detention.''\\

\textbf{Direct Challenge} & 
Confronting the opposing coder's perspective head-on, with clear language indicating a strong contrasting view. & 
Explicit Rejection~\cite{felton2009deliberation, bar2017stance, locher2004power},
Constructive Disagreement~\cite{felton2009deliberation, bar2017stance, locher2004power} & 
``\textbf{Coder 1’s} interpretation of the question as a check question \textbf{misrepresents} the essence of what constitutes a check question.''
\newline
``\textbf{Coder 1's} perspective \textbf{overlooks} this critical relationship, which justifies the labeling of Wilson's disease under digestive system diseases.'' \\

\textbf{Alternative Proposal}
& Suggesting an alternative viewpoint. 
& Alternative Proposal~\cite{bakhtin2010dialogic, parker2006public, chan2023chateval}, Reframing Perspective~\cite{bakhtin2010dialogic, parker2006public}
& ``Therefore, it might be better categorized as an \textbf{open-ended question [a third alternative option distinct from the views of Coder1 and Coder2]} instead of a clear check question.''
\newline
``Therefore, I propose labeling it as \textbf{undecidable case [a third alternative option distinct from the views of Coder1 and Coder2]} due to this ambiguity.'' \\

\textbf{Tentative Stance}
& Expressing an opinion with caution, acknowledging ambiguity or lack of certainty.
& Tentative Expression~\cite{felton2009deliberation, lu2011collaborative, angouri2010you, cheng2017anyone}
& ``Thus, it is \textbf{ambivalent}; it does not definitively confirm or refute the existence of guiding principles.''
\newline
``However, the \textbf{ambiguity} surrounding the intention behind "right?" makes it difficult to firmly categorize this utterance as either a check question or not.''
\newline
``The \textbf{lack of clear} legal standards or precedents regarding the threshold for inhumane treatment in this context leaves room for interpretation.'' \\

\textbf{Seeking Consensus}
& Showing willingness to align or compromise with others. 
& Mutual Agreement~\cite{teasley2008cognitive, diziol2010using, schwarz2018orchestrating, chan2023chateval, zhang2023exploring} 
& ``I \textbf{agree} with Coder 2's revised assessment that the case presents complexities that make it difficult to categorize definitively under Article 5.''
\newline
``I \textbf{concur} with Coder 1's conclusion that this utterance should be labeled as 0.'' \\

\textbf{Collaborative Reflection}
& Explicitly referencing how the interaction with the other coder’s viewpoint has led to a new understanding or shift in thought.
& Dual Perspective Recognition~\cite{felton2009deliberation, lu2011collaborative, angouri2010you, cheng2017anyone}
& ``\textbf{After considering Coder 1's perspective} that the utterance does not ask a check question and instead presents a rhetorical comparison, ...''
\newline
``\textbf{Upon reevaluating the concerns raised in Coder 1's position}, I acknowledge that hepatoblastoma, being a malignancy originating in the liver, is indeed related to the digestive system as the liver plays a critical role in digestion and metabolism.'' \\

\textbf{Redundant Restatement}
& Restating an earlier argument without introducing new information or distinct perspectives.
& Repetition~\cite{zakharov2021discourse}
& Coder 1: For the code ``Person'' to be applicable, there should be a sense of emotional resonance or significance linked to that person in relation to the music, which is absent here. 
\newline
Coder 2: ...
\newline
Coder 1: For the ``Person'' code to apply, there needs to be a layer of emotional resonance or significance that connects the person to the music, which is not evident in this instance. \\

\textbf{Explicit Contrast}
& Directly contrasting their own reasoning with the other coder’s point, highlighting the differences before affirming a stance.
& Arguments by Comparison~\cite{walton2008argumentation}
& ``\textbf{While} Coder 1 argues it \textbf{lacks the explicit structure of a check question [the other coder’s opinion]}, it does in fact \textbf{seek clarification about the action [their own opinion].}''
\newline
``The utterance `Could you clarify what you mean by `having the waves together'?' does not \textbf{ask a check question [the other coder’s opinion]}; \textbf{instead}, it \textbf{requests clarification [their own opinion].}'' \\

\textbf{Reflective Reconsideration}
& Indicating a review of prior discussion before concluding.
& Partial Concurrence~\cite{felton2009deliberation, lu2011collaborative, angouri2010you, cheng2017anyone}
& ``After reconsidering the earlier discussion ...''
\newline
``After reviewing both opinions, ...''
\newline
``After reflecting on the ongoing discussion, ...'' \\
\bottomrule
\caption{Comparison of Discursive Moves Between LLM Coders and Humans. This table summarizes the discursive moves observed in LLM coders' discussions to those reported for humans in existing research. The table also includes detailed descriptions and examples for each move.}
\label{tab:discursive_moves_llm} \\
\end{longtable}

We observed the discussion behaviors of LLM agents without additional fine-tuning or instruction about discursive moves and compared them with human discursive moves, providing a baseline that other researchers can expand and tailor for diverse applications.

Table~\ref{tab:discursive_moves_llm} contains the comparison between human discursive moves from the literature and the ones that are observed from AI coders’ discussions. The discursive moves that are absent from the AI coders but still used by humans are organized in the table in Appendix~\ref{app:discursive_moves_non_absent}.
LLMs replicate many of the discussion patterns of human coders, though key distinctions remain evident. For example, LLMs emulate basic patterns of interaction, such as maintaining stances, acknowledging others’ perspectives, reassessing evidence, explicitly contrasting ideas, expressing a tentative stance, and proposing alternatives. These behaviors highlight their capacity for logical and systematic reasoning. Additionally, LLMs are capable of supporting essential aspects of constructive discussions, such as fostering collaborative reflection and seeking consensus.

However, certain key aspects of human discursive behavior are absent in LLMs. Human discussants actively influence the flow and direction of discussions through dynamic interactions, such as raising questions to address uncertainties (e.g., clarifying ambiguities or challenging evidence), expanding on others' ideas, sharing personal experiences, and moderating discussions based on evolving conversational contexts. These actions reflect humans' ability to adjust their communication based on subtle shifts in the discussion and the reactions of other participants. In contrast, LLMs tend to operate within a more fixed framework, often missing the fluidity and adaptability required to respond to implicit cues or evolving conversational dynamics.

Moreover, LLMs do not replicate the emotional expressiveness inherent in human discourse. Human discussants often incorporate sarcasm and complaints, which can disrupt the focus of discussions or lead to unnecessary tension. This absence of emotionally charged or overly critical tones in LLMs can sometimes foster a more neutral and focused environment for logical reasoning. Instead, agents use softening language to foster a cordial atmosphere during the discussion. However, LLMs do share certain non-constructive discursive behaviors with humans, such as redundant restatement, which reiterates their earlier arguments without introducing new perspectives.

Meanwhile, humans use rhetorical strategies, such as analogies, to enhance persuasiveness and concretize abstract concepts. By comparison, while LLMs are proficient in generating well-structured and logical narratives, they frequently lack the creative subtlety needed to craft compelling narratives.

It is worth noting that LLMs engage in significantly less questioning compared to humans. While humans frequently ask clarifying or probing questions to refine understanding or actively explore the perspectives of other participants, LLMs tend to prioritize delivering coherent and direct answers over engaging in exploratory dialogue and mutual learning.

The differences between humans and LLMs are not limitations but complementary features that can be harnessed for mutual reinforcement in collaborative settings. Researchers can design human-AI collaborations by integrating the structured and efficient reasoning of LLMs with the creativity, emotional intelligence, and contextual awareness of humans. The discussion dataset we created enables the evaluation of LLMs' baseline discussion capabilities without fine-tuning, offering a foundation for guiding and benchmarking automated qualitative coding methodologies.
\section{Design Recommendations for Automated Qualitative Coding Systems}

Based on results from LLM experiments and data analysis, we propose several design recommendations for developers and designers of automated qualitative coding systems. The quantitative findings of experiments, including mean, standard deviation, minimum, and maximum values, are available in the Supplementary Materials. However, the primary aim of design recommendations is not to specify rigid thresholds but to offer a conceptual and methodological foundation upon which future research may further extend by focusing more directly on quantitative metrics.

\subsection{Design Recommendation for Designers 1: Provide users with guidance on preprocessing their datasets to optimize system performance.} 

One key recommendation is that designers should provide users with targeted guidance on how to preprocess their datasets to optimize system performance. This includes embedding features in the system that offer diagnostic feedback on dataset properties and provide actionable suggestions. For instance, systems can notify users when certain data characteristics are likely to result in reduced performance, such as when particular data for coding or code definitions may yield lower accuracy. In such cases, users can be advised to allocate additional scrutiny to the associated segments, or the system can generate preprocessing recommendations accordingly. 

The example suggestions may address codebook development such as encouraging concise codebook creation. Below are representative examples from the data analysis.

\begin{quote}
\textbf{<Example of Short Codebook>}\newline
\textit{To be labeled as “Article03”, the court case should violate the Article03. 
Article03: Prohibition of torture
No one shall be subjected to torture or to inhuman or degrading treatment or punishment.}

\textbf{<Example of Long Codebook>}\newline
\textit{To be labeled as “Article06”, the court case should violate the Article06.
Article06: Right to a fair trial
\newline1. In the determination of his civil rights and obligations or of any criminal charge against him, everyone is entitled to a fair and public hearing within a reasonable time by an independent and impartial tribunal established by law. Judgment shall be pronounced publicly but the press and public may be excluded from all or part of the trial in the interests of morals, public order or national security in a democratic society, where the interests of juveniles or the protection of the private life of the parties so require, or to the extent strictly necessary in the opinion of the court in special circumstances where publicity would prejudice the interests of justice.
\newline2. Everyone charged with a criminal offence shall be presumed innocent until proved guilty according to law.
\newline3. Everyone charged with a criminal offence has the following minimum rights:
… \\}
\end{quote}

Another recommendation for the codebook can include avoiding repetitive or circular code definitions, which result in a high similarity between words within a codebook, and instead employing multiple, distinct descriptions that have a low similarity between words within a codebook.

\begin{quote}
\textbf{<Example of Low Similarity Codebook>}\newline
\textit{To be labeled as “Person”, the respondent should associate the track with a person such as a family member, a friend, the loss of a person, a group of people, or not being with a person or being alone.}

\textbf{<Example of High Similarity Codebook>}\newline
\textit{To be labeled as “Negative Evaluations”, the respondent should make a negative evaluation about the track or an aspect of the track. \\}
\end{quote}

Users may also be prompted to split data for coding into chunks of one to two sentences.

\begin{quote}
\textbf{<Example of Short Excerpt>}\newline
\textit{So do you know what virtual reality is?}

\textbf{<Example of Long Excerpt>}\newline
\textit{This is a great question because it really gets to the difference between some of our different tools in our machine learning tool belt in addressing problems like this. So if we were to use a supervised learning classic classification approach, a person would need to think about those features and creatively come up with them in approach we call the kitchen sink approach, which is just try everything you can possibly think of and see what works.}
\end{quote}

The qualitative data for coding should then be organized around coherent or limited topics using NLP techniques such as TextTiling. 

\begin{quote}
\textbf{<Example of Low Similarity Excerpt>}\newline
\textit{Slightly like slipknot. It reminds me of the game ``need for speed underground 2'' a little. The repetitive riffs make me think of grinding the same missions or tasks at work or in certain games. There is a certain anger, or energy brought by the vocals that makes me knuckle down, but the guitar is mostly happy and upbeat. I went through a phase of listening to the GazettE, but I still think it’s good. Energetic and upbeat. I don't get too distracted by the lyrics as I don't speak Japanese. It helps me focus and power through. Nope, it's just a song I quite like.}

\textbf{<Example of High Similarity Excerpt>}\newline
\textit{A sense of calmness and serenity. Because it puts my life into perspective when I feel anxious or worried about something. Because it calms me down and makes me think of only positive things, all negative things just fall away. Yes, it's a Christian track, so when I play it, it makes me think of God and his promises, so it gives me a perspective and reminder that I am loved.}
\end{quote}

Furthermore, designers should encourage users to match the degree of specialization in their codebooks with that in their datasets. 

\subsection{Design Recommendation for Designers 2: Allow users to leverage (1) the ``Undecidable'' label and (2) the conflicting label as a performance indicator to predict potential accuracy improvements resulting from AI agent discussions.}

Beyond preprocessing, systems should enable users to understand and use performance signals. In particular, the “Undecidable” and conflicting labels should be made interpretable as indicators of potential accuracy gains via AI agent discussions. For example, users can be guided to interpret the number of uncertainty prior to AI agent discussions as a potentially positive sign, and the number of uncertain labels after discussion as a negative indicator of limited benefit from the process. Likewise, persistent conflict between agents following discussions should be treated as a potential signal of limited effects of AI agent discussions. Incorporating such diagnostic tools allows users to develop informed mental models of the system’s behavior and adjust their expectations.

\subsection{Design Recommendation for Designers 3: Allow users to leverage the level of controversy within AI agent discussions as a performance indicator for evaluating the expected accuracy of discussion-prompted coding.} 

Another major recommendation addresses the value of the degree of agent disagreement. While disagreements might be intuitively understood as limitations, our findings suggest that they are often more informative than their resolution. Designers should reconceptualize agent disagreement during coding discussions not as a failure but as a signal of interpretive richness. Our analysis indicates that prolonged or unresolved disagreements between agents often yield clarification of coding criteria and foster a more nuanced understanding of edge cases. In some cases, the resolution of disagreements may correspond with low information gain, suggesting that users should capitalize on the presence and structure of disagreement rather than its resolution alone. Specifically, in gauging the expected accuracy of discussion-prompted coding, designers can guide users to interpret high volumes of conversational exchange between agents as a positive sign while viewing high proportions of resolved or even correctly resolved conflicts as potential negative indicators. Below are representative examples from the data analysis.

\begin{quotation}
\textbf{<Example of Controversial Discussion>}\newline
...\newline
\textit{Coder 1: I appreciate Coder 2's interpretation that ``what about negative harmony'' could serve as a discussion prompt. However, the nature of the phrase ``what about'' often leans towards rhetorical questioning rather than a genuine inquiry seeking specific information. ...
\newline
Coder 2: After considering Coder 1's arguments, I maintain that the phrase ``what about negative harmony'' does indeed present a question. While Coder 1 suggests it functions more as a prompt, I argue that it invites exploration of a specific topic, which aligns with the essence of D02. ...
\newline
Coder 1: I appreciate Coder 2's perspective that ``what about negative harmony'' could be seen as an invitation for discussion; however, I still contend that it lacks the qualities of a genuine inquiry. ...
\newline
Coder 2: I understand Coder 1's view regarding the rhetorical nature of ``what about negative harmony,'' yet I continue to assert that it functions as an inquiry. ...}

\textbf{<Example of Non-Controversial Discussion>}\newline
...\newline
\textit{Coder 1: While I understand Coder 2's perspective that the excerpt discusses potential and challenges, I believe it still conveys specific information about opportunities in synthetic worlds and the limitations of current simulation technologies. ...
\newline
Coder 2: Upon reevaluating my position and considering Coder 1's reasoning, I recognize that the utterance does indeed convey substantial information about the opportunities and limitations involved in creating synthetic fantasy worlds. Phrases like ``there's tons of opportunities for the synthetic case'' and ``we still don't do people well simulated'' provide clear insights into the current state of technology in this area. Therefore, I agree with Coder 1 that this aligns with the requirements for D09, as it effectively informs the listener about specific aspects of the topic.}
\end{quotation}

\subsection{Design Recommendation for Designers 4: Adjust AI agents' discussion style based on the intended objectives of the coding task.}
A final recommendation focuses on discussion dynamics and the adjustment of agent behaviors according to the objectives of the coding task. Designers may consider providing users with the option to activate a discussion mode that deliberately introduces controversy in tasks where ambiguity is prevalent in the dataset. 
When a label is undecidable, confidence is low, or prior performance has been unreliable, the system can shift into a debate-enhancing mode that fosters disagreement, prompting agents to adopt explicitly opposing perspectives. Our findings suggest that introducing structured opposition within agent discussion improves the accuracy of discussion-prompted coding.

\begin{quotation}
\textbf{<Example of Discussion with Competing Mode>}\newline
\textit{Upon further reflection, I maintain my stance that the excerpt contains positive evaluations. … While Coder 1 argues these are not explicit evaluations, they do reflect a positive judgment about the track and its context, fulfilling the criteria for Positive Evaluations as outlined in the description.}

\textbf{<Example of Discussion with Compromising Mode>}\newline
\textit{While I understand Coder 2's perspective regarding closed head injury being a significant condition, I believe we are dealing with an undecidable case here. … Therefore, based on the provided data, it is difficult to definitively categorize it as either present or absent under this specific label [, suggesting the Undecidable label as an alternative].}

\textbf{<Example of Discussion with Collaborating Mode>}\newline
\textit{Upon reevaluating Coder 1's reasoning, I recognize that the term ``nostalgic'' serves as a descriptive adjective that evokes specific emotional qualities associated with the track. … Therefore, I change my initial assessment to agree with Coder 1.}
\end{quotation}
\section{Discussion}

This section examines the key findings of our work, acknowledges the study's limitations, and outlines directions for future research.

\subsection{Summary of Key Findings and Implications}

Our study reveals that LLM “coders” can engage in a structured dialogue to perform qualitative coding, with performance influenced by specific dataset characteristics and discussion dynamics. First, we found that coding accuracy is shaped by dataset’s length and semantic similarity, and the gap of degree of specialization between codebook and data for coding. These findings highlighted that preprocessing codebooks and qualitative data for coding can boost AI coding performance. 

Second, our results show that the frequency of “Undecidable” label and conflict between AI agents can be utilized as a performance indicator. These signals assist users in calibrating their expectations of automated systems. We discovered that instances of disagreement and debate between the AI coders can be a positive signal for high accuracy of coding. Rather than viewing disagreement as a failure, our study indicates that when the two AI coders disagreed on how to code a segment, the conversation often increased interpretive depth. 

The AI’s convergence is a double-edged sword: on one hand, it led to consistent consensus in most cases, but on the other hand, we suspect it might sometimes mask uncertainty. For instance, in a few discussions we examined, one agent proposed a code interpretation that the other accepted quickly. Such instances resulted in a consensus label that both agents agreed on, yet the underlying ambiguity of the data might not have been fully resolved, leading to an incorrect labeling outcome. This behavior underscores a design consideration on how to interpret automated outcomes: simply because two AI agents agree after a short discussion does not guarantee the decision was correct or well-founded. Furthermore, prolonged exchanges were associated with a more nuanced understanding of edge cases, whereas quick resolutions sometimes signaled that less information was gained through discussions. Together, these findings ground our discussion in concrete evidence from the experiment, moving beyond abstract statements to actionable recommendations for design.

Third, our qualitative analysis of LLM discourse behavior versus human coder discussions uncovered both parallels and gaps. On one hand, the AI agents replicated many constructive discussion patterns observed in human teams. They maintained or defended their initial stances, acknowledged each other’s perspectives, contrasted differing interpretations explicitly, reassessed evidence, and proposed alternatives. These behaviors suggest that LLMs are capable of logical, systematic reasoning. On the other hand, we identified several human discussion strategies that were absent or weaker in the AI conversations. For example, human coders often interject questions to clarify uncertainties, challenge assumptions, or seek additional evidence; in our study, the LLM agents did not engage in clarifying or probing questions. Humans also dynamically adjust the discussion flow, moderate the direction of debate, share personal anecdotes, and resort to figurative language and rhetorical strategies to enhance engagement and clarity. In contrast, the AI discussions followed a more fixed pattern, not fully adapting to nuanced conversational cues or shifting strategy mid-discussion. Emotional or stylistic nuances present in human dialogue, such as sarcasm or frustration, were essentially absent from the AI’s exchanges. This lack of emotional reactivity in LLMs kept the tone neutral and focused, potentially creating a more consistently cordial environment for deliberation. Meanwhile, the AI agents did exhibit some less productive tendencies akin to humans, such as repetitive restatement of points without adding new information. In summary, LLM-based coders can mirror basic human reasoning moves but miss the adaptive and inquisitive depth of human interaction, suggesting that while LLMs excel at structured reasoning, they may fall short in demonstrating the more affective and socially attuned aspects of human discourse without additional fine-tuning.
 
\subsection{Limitations}

While our study provides encouraging evidence for LLM-based qualitative coding, several limitations temper our conclusions. First, the AI model we used was not fine-tuned for qualitative coding tasks or any domain-specific data. All agent behaviors were elicited through prompting alone. This choice allowed us to examine out-of-the-box capabilities, but it also means performance might be improved with customization. A fine-tuned model or one explicitly instructed in effective discussion strategies could produce different (potentially better) outcomes.

Second, the evaluation of coding accuracy and consensus has inherent challenges. We treated the original human-coded dataset labels or expert consensus as the “ground truth” for accuracy calculations. However, in qualitative research, coder disagreements can sometimes reflect multiple valid interpretations rather than clear-cut errors. Our use of accuracy as a metric, while necessary for quantitative analysis, oversimplifies this nuance. An AI-assigned code different from the ground truth was counted as incorrect, even if it might be a justifiable interpretation in context. 

Third, the rapid pace of LLM development introduces another layer of complexity. New releases of LLMs may significantly alter the dynamics of agent discussions, potentially changing both the quality and style of coding collaboration. This evolving baseline makes it challenging to draw stable conclusions from experiments tied to a particular model version, and highlights the need for ongoing evaluation as the technology advances. Meanwhile, although our experiments were tied to GPT-4o-mini, the underlying dynamics — such as the tendency for accuracy gains to emerge from prolonged debate rather than rapid consensus, with controversy about borderline cases clarifying inclusion and exclusion criteria for codes and — represent model-agnostic mechanisms. These dynamics are likely to persist across LLM families, since they stem from the interactional structure of multi-agent debate rather than from model-specific capabilities. At the same time, given the rapid evolution of LLMs, we frame our findings less as static benchmarks and more as design-oriented insights that persist across model generations.

Finally, our study setting was a simulated one: AI agents discussed with each other, and we compared their behavior to documented human coding discussions from literature. We did not conduct a user study of human analysts actually working with these AI agents. As a result, we have not directly measured human researchers’ trust, satisfaction, or workflow integration when partnering with AI coders. These human factors are crucial for real-world adoption and were beyond our current scope. The absence of direct human-in-the-loop experimentation is a limitation that we acknowledge and aim to address in future work.

\subsection{Theoretical Implications: Rethinking Reliability and Agreement}

Beyond practical design, our work invites HCI and computational social science researchers to reconsider some theoretical assumptions in qualitative research when one of the coders is an AI. In traditional qualitative coding, intercoder reliability (e.g., Cohen’s Kappa) between human coders serves as a check on consistency and code clarity. But what does reliability mean when a coder is an AI? We found that measuring agreement between two AI coders was useful for our analysis, yet it differs from human scenarios. 

For one, AI coders (especially if they are instances of the same LLM model) may have correlated errors or shared biases, violating the assumption of independent judgments. In our study, both AI coders were based on the same underlying model architecture and were exposed to the same training data distribution, which is unlike two independent human coders with distinct backgrounds. This means a high agreement between AI coders might not guarantee the validity of the coding — they could be consistently wrong in the same way, implying that reliability statistics need reinterpretation

Inter-rater reliability with an AI partner could be overly optimistic (if the AI’s decisions are rubber-stamped by a human or vice versa) or overly pessimistic (if the AI systematically challenges a human coder on edge cases that humans might ordinarily skip as noise). One theoretical extension suggested by our work is to develop new metrics that account for the discussion process. Instead of just percent agreement on final codes, one could measure the proportion of cases that required discussion to reach agreement, or the stability of the AI’s decisions before and after debate. These measures might capture the nuance of AI coding interactions better than a single agreement coefficient. 

Additionally, we propose rethinking the role of codebooks and coding guidelines for AIs. When an AI is involved, perhaps the codebook should be written not just for human interpretability but also optimized for machine interpretability. 

Another implication is how we define consensus. In human teams, consensus is often the end of discussion; in human–AI teams, we might deliberately choose not to fully automate consensus. It could be acceptable, or even desirable, to have some lasting disagreements, if those indicate areas for further human analysis. Theorists and practitioners should be open to the idea that a discord between AI coder or between a human and an AI coder could spark deeper insights. 

Meanwhile, the distinction between what is ‘reasonable practice’ and what is ‘inherent to qualitative analysis’ illuminates a question: is qualitative coding reducible to a standardized analytic procedure, or is it constitutively interpretive? We take the latter view. The presence of ambiguity and disagreement is not a byproduct of methodological insufficiency of imperfect method but a constitutive condition of qualitative knowledge production. Our contribution is to formalize its role in computational pipelines. By operationalizing disagreement and undecidability, we reposition what qualitative research has treated as unavoidable and generative into a measurable feature that informs the design of automated systems.

In summary, our research calls for an expanded notion of reliability and consensus in coding when integrating AI – one that values the process of reaching agreement (or not reaching it) as much as the final agreement itself.

\subsection{Design Implications: LLM-Based Coding Systems}

Our findings suggest several important design implications for systems that incorporate LLMs into qualitative coding workflows. One key design recommendation is to offer tunable AI behavior modes, specifically a toggle between a “strict” mode and a “debate” mode. In the debate-enhancing mode, the system could detect low-confidence scenarios — such as undecidable cases or potential coding conflicts — and then trigger divergent perspectives among AI agents. By encouraging structured disagreement or devil’s advocacy, the system can prompt deeper analytical engagement and help surface nuanced considerations that may be lost in overly harmonious exchanges. Our results indicate that such structured debate is especially beneficial for ambiguous or complex coding situations, improving the overall accuracy and depth of analysis. On the other hand, for routine or well-defined coding tasks, a concise exchange may be preferable. In these scenarios, a “strict” mode could streamline interactions. This dynamic control over the discussion mode offers alignment with the user’s task goals. For example, a complicated qualitative analysis might benefit from competitive and discursive outputs clarifying challenging cases, whereas quick coding for a simple survey might prefer efficiency. 

While our study focused on AI–AI coder pairs, a future direction is the integration of human coders directly into the discussion loop. A human-in-the-loop design could better align the coding process to users' goals and intent by allowing human users to guide and challenge AI-generated codes in real time.

Qualitative coding has long been theorized as more than a procedural act of labeling; it is an interpretive practice in which researchers mediate between data, theory, and their own positionality. Reflexivity — acknowledging how one’s standpoint and assumptions shape interpretation — is central to ensuring validity in this tradition. While our study demonstrates that LLMs reproduce discursive moves reminiscent of human coders (e.g., stance maintenance, acknowledgment, tentative stances), these moves operate as surface-level simulations of interpretive dialogue. They do not embody the meta-cognitive work of interrogating one’s assumptions, nor the positional disclosure that grounds qualitative validity. As a result, AI “agreement” may project an illusion of rigor while bypassing the epistemic labor that human researchers contribute through reflexive engagement.

For automated qualitative coding, this implies that validity must be reconceptualized as a co-constructed property of human – AI interaction. AI coders can highlight ambiguity, surface discursive conflicts, and generate systematic consistency, but reflexivity remains the province of human researchers. Rather than viewing LLM consensus as a substitute for interpretive rigor, we argue that its true value lies in provoking reflexive engagement: Why did the models disagree? What kinds of ambiguity do their debates reveal? How does a researcher’s theoretical stance shape the interpretation of such ambiguity? Designing systems that scaffold these questions — by embedding reflexive checkpoints, prompting positional notes, and visualizing where AI outputs diverge from human interpretive expectations—can transform automation from a tool of decontextualization into a partner in reflexive inquiry. In this way, the uniquely human aspects of coding are not displaced but re-centered, with AI disagreements serving as catalysts for deeper interpretive practice.

We also encourage longitudinal studies and domain-specific deployments of LLM-based coding to see how our findings hold in real-world practice. For instance, deploying an AI coding assistant in a social science research lab for several months could uncover how human coders adapt their strategies when an AI is part of the team — do they come to trust the AI for certain codes more than others? Does the presence of an AI change how teams construct their codebook? Different domains (e.g., healthcare interviews vs. Twitter data vs. HCI user study feedback) may present unique challenges for LLM coders. Future work should examine a variety of contexts, potentially leading to specialized design guidelines.

Several directions emerge from our study that could extend and deepen this line of inquiry:

\begin{itemize}
  \item \textbf{Richer analysis dimensions:} Future studies could vary the number of AI agents, assign them distinct personas or expertise areas, and experiment with multiple iterations of discussion to observe how arguments evolve or converge over time.
  \item \textbf{Tracking dynamics of iterative discussions:} Our study involved a single iteration of discussion. Meanwhile, investigating how the structure and tone of AI discussions change across multiple iterations could offer insights into discussion quality and decision-making patterns.
  \item \textbf{Customized discussion behaviors:} Based on observed behaviors, systems could adapt the agents’ conversational styles or debate strategies to match the user's preferences or the coding context.
  \item \textbf{Human-in-the-loop pipeline design:} Designing workflows where human coders participate proactively — adding feedback to AI debates, inserting their own counterpoints, or steering the discussion toward specific frames — can make the process better align with the user’s intent and goals. Future tools could allow humans to reflect their opinions within AI debates, not just through final selections but as part of the ongoing discussion, enabling richer co-creation and mutual learning between humans and AI.
\end{itemize}

\section{Conclusion}

Our work contributes a foundational step toward understanding and improving automated qualitative coding using LLMs. By simulating coder discussions, examining the impact of data characteristics, and analyzing AI behaviors qualitatively, we provide a nuanced account of the strengths and limitations of LLMs in replicating human coding dynamics. Our findings emphasize that disagreement and ambiguity are not merely challenges but opportunities for deeper insight, and they should be strategically harnessed in system design. From these findings, we propose a set of design recommendations for designers: maintain concise and diverse codebooks, structure excerpts around coherent topics, use coding uncertainty like ``Undecidable'' labels as diagnostic tools for coding accuracy, and tune agent discussion style to suit task goals.


\bibliographystyle{ACM-Reference-Format}
\bibliography{sample-base}

\appendix


\section{Comparison of Discursive Moves Between LLM Coders and Humans}
\label{app:discursive_moves_non_absent}

\begin{longtable}{
  p{6cm}
  p{7cm}
}
\caption{Comparison of Discursive Moves Between LLM Coders and Humans. This table shows the discursive moves used by humans in the literature, which did not appear in the discussions of LLM coders.}
\label{tab:discursive_moves_literature} \\

\toprule
\textbf{Category of Discursive Moves} & 
\textbf{Discursive Move of Humans in Literature} \\
\midrule
\endfirsthead

\toprule
\textbf{Category of Discursive Moves} & 
\textbf{Discursive Move of Humans in Literature} \\
\midrule
\endhead

Moderating & Moderation~\cite{grice19901975, wise2011analyzing, bakhtin2010dialogic} \\
Illustrating & Example and Illustration~\cite{walton2008argumentation, poppel2020study}\\ & Metaphor and Analogy~\cite{walton2008argumentation, poppel2020study} \\
Expansive & Idea Expansion~\cite{zakharov2021discourse, walton2008argumentation, poppel2020study} \\
Personal Story & Personal Narrative~\cite{zakharov2021discourse, walton2008argumentation, poppel2020study} \\
Questioning & Clarification Request~\cite{grice19901975, bakhtin2010dialogic} \\
 & Questioning Evidence~\cite{zakharov2021discourse, walton2008argumentation, poppel2020study} \\
 & Answer~\cite{zakharov2021discourse, walton2008argumentation, poppel2020study} \\
 & Critical Question~\cite{zakharov2021discourse, walton2008argumentation, poppel2020study} \\
Aggressive & Confrontational Expression~\cite{zakharov2021discourse, walton2008argumentation, poppel2020study} \\
 & Dismissive Humor~\cite{zakharov2021discourse, walton2008argumentation, poppel2020study} \\
 & Sarcasm~\cite{zakharov2021discourse, walton2008argumentation, poppel2020study} \\
 & Reframing Critique~\cite{zakharov2021discourse, walton2008argumentation, poppel2020study} \\
Disruptive & Relevance Critique~\cite{zakharov2021discourse, walton2008argumentation, poppel2020study} \\
 & Nitpicking~\cite{zakharov2021discourse, walton2008argumentation, poppel2020study} \\
 & Complaint~\cite{zakharov2021discourse, walton2008argumentation, poppel2020study} \\
 & Disruptive Behavior~\cite{zakharov2021discourse, walton2008argumentation, poppel2020study} \\
 & Topic Derailment~\cite{zakharov2021discourse, walton2008argumentation, poppel2020study} \\
 & Non-reasoned Disagreement~\cite{zakharov2021discourse, walton2008argumentation, poppel2020study} \\

\bottomrule
\end{longtable}

\end{document}